\documentclass[showpacs,amssymb,singlecolumn,floatfix,pre,aps,notitlepage,unsortedaddress]{revtex4-1}

\usepackage[utf8]{inputenc}

\usepackage{amssymb,amsfonts,amsmath,bm}
\usepackage{graphics,graphicx}
\usepackage{xcolor}
\usepackage{hyperref}
\usepackage{comment}

\begin{document}

\title{Singular relaxation of a random walk in a box with a Metropolis Monte Carlo
dynamics}
\author{Alexei D. Chepelianskii}
\affiliation{Universit\'e Paris-Saclay, CNRS, Laboratoire de Physique des Solides, 91405 Orsay, France.}
\author{Satya N. Majumdar}
\affiliation{Universit\'e Paris-Saclay, CNRS, LPTMS, 91405 Orsay, France.}
\author{Hendrik Schawe}
\affiliation{LPTM, UMR 8089, CY Cergy Paris Universit\'e, CNRS, 95000 Cergy, France.}
\author{Emmanuel Trizac}
\affiliation{Universit\'e Paris-Saclay, CNRS, LPTMS, 91405 Orsay, France.}

\date{\today}

\begin{abstract}
We study analytically the relaxation eigenmodes of a simple Monte Carlo algorithm, corresponding to a particle in a box which moves by uniform random jumps. 
Moves outside of the box are rejected. At long times, the system approaches the equilibrium probability density,
which is uniform inside the box.
We show that the relaxation towards this equilibrium is unusual: 
for a jump length comparable to the size of the box, the number of relaxation eigenmodes can be surprisingly small, one or two. We provide a complete analytic description of the transition between these two regimes. When only a single relaxation eigenmode is present, a suitable choice of the symmetry of the initial conditions gives a localizing decay to equilibrium.
In this case, the deviation from equilibrium concentrates at the edges of the box where the rejection probability is maximal. 
Finally, in addition to the relaxation analysis of the master equation, we also describe the full eigen-spectrum of the master equation including its sub-leading eigen-modes.
\end{abstract}
\maketitle

\section{Introduction}

 Random sampling techniques of the Monte Carlo type allow to solve both deterministic and stochastic problems. While the original Metropolis–Hastings algorithm was introduced in the context of classical many-body statistical physics \cite{Metropolis_Ulam}, generalisations of this approach are now employed in a wide range of topics from basic sciences to planning and forecast in economy or epidemiology \cite{Binder_book,FrenkelSmith,mode2011applications,bolhuis2002transition,becca2017quantum,bishop2006pattern,Shaebani2020,Rubinstein,Glasserman2003,Gilks}.
Markov Chain Monte Carlo techniques \cite{Roberts,NewmanBarkema} provide a class of simple yet powerful algorithms, where the target probability distribution is obtained as a sequence of samples from a random walk with appropriate transition probabilities. In Metropolis-Hastings algorithms, some of the attempted moves can be rejected in which case the state of the system remains unchanged during an iteration step. The probability of rejection is set to ensure detailed balance \cite{Binder_book,NewmanBarkema,FrenkelSmith}.  The ratio between rejected and accepted moves impacts the convergence rate of the algorithm and depends on the probability distribution of the attempted jumps. A widely accepted empirical rule is that the acceptance probability for the attempted moves be close to 50\% 
for parameters ensuring optimal convergence \cite{Krauth,Allen,FrenkelSmith,Talbot}. 
A number of mathematical results have been proven for the relaxation rates of the Metropolis algorithm, however those mainly hold close to the diffusive limit of small jumps \cite{Dey2019,Jourdain2015,Peskun1973,Goldilocks} where powerful mathematical techniques based on micro-local analysis have been developed \cite{Diaconis2009,Diaconis2011,Diaconis2012} allowing to obtain exact results on the relaxation rate.

The evolution of the probability distribution with the number of algorithmic steps is in general described by a linear Master equation. When detailed balance is enforced, the target probability distribution can be shown to be an invariant eigenvector of the Master equation. While general theorems then guarantee convergence to the desired steady state \cite{Wasserman, Levin},
little is known about the complete eigenspectrum of Master equations and its eigenvectors and the available results are mainly numerical \cite{Diaconis,Diaconis_1}. For Markov chains on a finite number of states, the number of eigenvalues is given by the number of states and the relaxation rate is determined by the eigenvalue with the largest absolute value $|\lambda|<1$ ($\lambda=1$ being associated to equilibrium/steady state). However, Markov chain Monte Carlo  algorithms often involve continuous variables (for example particle coordinates in configuration space), and in such cases there exists no general theorem on the structure of the eigenspectrum. In \cite{MC2022}, we showed that the balance between acceptance and rejection for continuous variables can lead to a transition in between two regimes. In addition to the standard situation where the relaxation rate is ruled by the leading eigenvalue of the Master equation, there appears a new regime, as a function of the typical jump length,  where the relaxation is instead governed by the maximal possible rejection probability. In this regime, the difference to the expected steady state of the algorithm progressively localises around the points where the rejection probability is maximum, in a way somewhat reminiscent of the zero-temperature Monte Carlo dynamics \cite{Chepelianskii_2021}. This analogy led us to describe the transition between the two regimes as a localization transition, yet without a complete analytical understanding. In this article, we analyse a simple continuous Markov process for which the eigenspectrum and eigenvectors can be computed analytically, allowing to describe analytically the full spectrum of eigenvalues/eigenvectors and thus to investigate the localization transition in depth. We define in section \ref{sec:setting} the random walk confined in a box that underlies the Monte Carlo dynamics considered,
and discuss the generic relaxation behaviour that ensues. A key quantity is the amplitude of the random jumps attempted, and depending
on its value, different regimes govern the long-time evolution of the system. They are described in section
\ref{sec:large} for large jumps, where the number of relaxation eigenmodes is smallest, and then
for smaller jumps in section \ref{sec:a2p3to1}. An original scenario for relaxation ensues,
where the difference between the long-time and the equilibrium solutions
becomes ``pinned'' at the edges of the box.
In section \ref{sec:summary}, we summarize 
our spectral findings, and their connections to relaxation. Conclusions are presented in section \ref{sec:conclusion}. In the appendix we present a detailed comparison between the analytical calculations and numerical simulations investigating the relaxation of the Metropolis algorithm through direct numerical experiments or through numerical simulations of the master equation with excellent agreement between analytical results and numerical simulations.

\section{The setup: Master equation and relaxation dynamics}
\label{sec:setting}

\subsection{The Metropolis rule}

\begin{figure}
\includegraphics[width=0.6\textwidth]{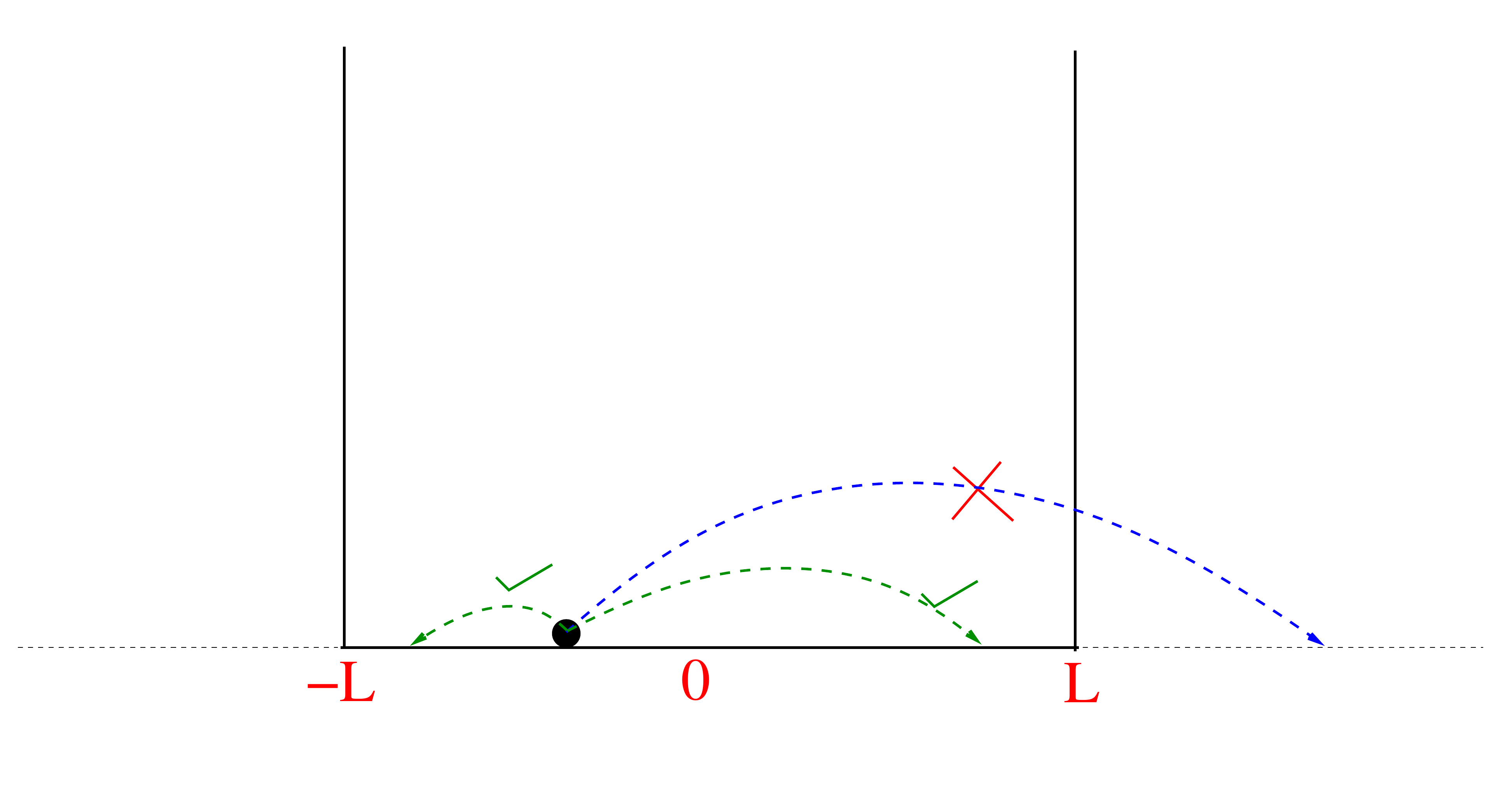}
\caption{A particle (shown by the filled disk) in a box $[-L,L]$  attempts to jump at each discrete time by a random displacement $\eta$ drawn independently 
from a symmetric and continuous distribution $w(\eta)$. 
If the attempted jump takes the particle outside the box $[-L,L]$, the move is rejected (as shown
schematically by the blue dotted line with a red cross forbidding it) and the particle stays at its current
position. If however the
jump takes the particle to a new position within the box $[-L,L]$, the jump is allowed and the
particle moves to the new position (allowed displacements shown by the dotted green lines).}
\label{box1.fig}
\end{figure}

In this paper, we focus on a simple random walk confined in a box $[-L,L]$:
at each discrete time step, the walker attempts a jump from position $x_{n-1}$
to 
$x_n = x_{n-1} +\eta_n$, where $\eta_n$ is a random variable drawn from the
distribution $w\eta)$.
The jump $\eta_n$ is accepted provided 
$-L \leq x_n \leq L$, see Fig. \ref{box1.fig}.
This process leads to the following Master equation for the probability distribution function (PDF) $P_n(x)$ of the random walker:
\begin{equation}
P_n(x)= \int_{-L}^L P_{n-1}(x') \,w(x-x')\, dx' + R(x) P_{n-1}(x) .
\label{fp.1}
\end{equation}
The first term on the right-hand side describes the
net probability flux to the position $x$ at step $n$ from other positions $x'$ at step $(n-1)$.
The second term describes the probability that the particle is at position $x$ at step $(n-1)$ and
does not move from this position at the next step. The term:
\begin{align}
R(x) = 1 - \int_{-L}^L w(x'-x)\,dx' 
\end{align}
describes the probability that the attempted move from $x$ is rejected. The function  $P_0(x)$ gives the initial PDF of the random  walker and sets the initial condition for the Master equation.

We can  check some general properties of this Master equation.
By integrating Eq. (\ref{fp.1}) over $x$, it can be verified
that the total probability at each step $n$ is conserved.
Another general check is that $P_\infty(x)=1/(2L)$ for $x\in [-L,L]$ is a fixed point
solution of Eq. (\ref{fp.1}), irrespective of the jump PDF $w(\eta = x-x')$. 
Thus at long times, the system will approach this fixed point
as its stationary `equilibrium' solution. Since the jump distribution function is symmetric,  equation~(\ref{fp.1}) is directly self-adjoint and will have real eigenvalues in $(-1,1]$ 
(interestingly, this property is also true for a general $U(x)$ due to detailed balance \cite{Risken,MC2022}).

\subsection{The convergence rate and long-time evolution}

An information of central interest deals with the speed with which 
$\delta P_n \equiv P_n-P_\infty$ goes to zero, as time $n$ increases.
The convergence rate can be defined from the large time limit of the deviation from equilibrium of some observable ${\cal O}(x)$:
\begin{equation}
\log\Lambda  \,=\, \max_{\{{\cal O}(x),P_0(x)\}}  \lim_{n \to \infty} \frac{1}{n} \log \left| \int {\cal O}(x) \delta P_n(x) dx \right| 
\label{eq:Lambda_def}
\end{equation}
where the maximum is taken over all possible functions ${\cal O}(x)$ and initial distributions $P_0(x)$. If $\Lambda < 1$, the probability distribution $P_n(x)$ converges exponentially fast to the equilibrium distribution for large $n$, i.e., $|P_n(x)- P_\infty(x)| \propto \Lambda^n \propto e^{-n/\tau}$ where
$\tau = - (\log \Lambda)^{-1} $
denotes the decorrelation time (in number of Monte Carlo algorithmic steps).
The convergence rate $-\log(\Lambda) >0$ is the figure of merit of the algorithm; the smaller the $\Lambda$, the larger the relaxation rate. 
If the relaxation of the Master equation~(\ref{fp.1}) is governed by its eigenspectrum, 
$\Lambda = \max |\lambda|$ where the maximum is taken over all the eigenvalues $\lambda \ne 1$ of the Master equation. 
On the other hand, the algorithm relaxation rate can also be limited by rejection events: $\Lambda = \max_{x \in [-L,L]} R(x)$. 
For the box potential, the rejection probability is maximum at the boundary $x = \pm L$, where the rejection probability is $1/2$, for symmetric $w(\eta)$. This leads to the general bound
\begin{align}
\Lambda \ge 1/2 ,
\label{uniformlimit}
\end{align}
which sets an upper limit to the convergence rate. One of the principal goals 
of this work is to study how $\lambda$ depends on the typical jump length encoded
in $w(\eta)$.

It is possible to construct a formal singular eigenfunction with an eigenvalue given by the rejection probability $R(x)$ for any point $x$. Let us consider $\Psi_{R(x)}$ with $\Psi_{R(x)}(y) = 1$ only if $y = x$ and  $\Psi_{R(x)}(y) = 0$ for all other $y \ne x$. Substitution into the Master equation \eqref{fp.1} shows that $\Psi_{R(x)}$ is an eigenfunctions with $\lambda = R(x)$. Such solutions are zero almost everywhere. 
We nevertheless show below that they add a new term to the eigenvector expansion, which is not present for the Master equation of a Markov chain on a discrete set. To highlight the absence of smooth eigenfunctions, we will call $R([-L,L])$  the singular continuum of eigenvalues.

Below, by explicit solution of the Master equation for the box problem, we show that to account for the two possible relaxation scenarios, the eigenfunction expansion of the probability distribution reads
\begin{equation}
   P_n(x) \, =\, \sum_{\lambda \in \{\lambda_{0} \ldots \lambda_{\cal N} \}} {\cal A}_\lambda \,{\cal P}_\lambda(x) \, \lambda^n  \,+\, {\cal L}_n(x) ,
    \label{eq:eigen_discrete_continuous}  
\end{equation}
where ${\cal L}_n(x)$ describes  the contribution from the singular continuum and the amplitudes ${\cal A}_\lambda$ are fixed by the overlap of the initial distribution $P_0(x)$ with the eigenfunctions of the Master equation.
Here, the discrete summation runs over a finite (and possibly small) number of ${\cal N}+1$ terms,
with ${\cal N} \geq 0$. The eigenvalues are ordered such that $\lambda_0 \geq \lambda_1 \geq \lambda_2\ldots$\
with $\lambda_0=1$ corresponding to the equilibrium solution.
The remaining term ${\cal L}_n(x)$ localizes at large times $n\to \infty$ around a finite number of points $x_l$ where the rejection 
rate $R(x)$ in \eqref{fp.1} is maximal: $\lim_{n \rightarrow \infty} {\cal L}_n(x)/{\cal L}_n(x_l) = 0$ for any $x \ne x_l$. In this sense the term ${\cal L}_n$ asymptotically converges to a function which is zero almost everywhere, like the  eigenfunctions of the singular continuum. 
The case ${\cal N} = 0$ corresponds to the localizing case where only the equilibrium term $\lambda_0=1$ is present in the expansion and relaxation is entirely governed by the ${\cal L}_n$ term. We already derived in \cite{MC2022} an explicit analytical expression for the asymptotics governed by ${\cal L}_n$, for a problem with ${\cal N} = 0$; however, it was not possible to obtain explicit solutions for parameters with different values of ${\cal N}$. In the present article, we derive exact solutions for a wider range of parameters, allowing to describe analytically the change of behaviour between ${\cal N} = 1$ and ${\cal N} = 2$.

\subsection{Spectral problem for the box confinement}

To make analytic progress possible,
we consider in the remainder the case of a uniform jump distribution, as alluded to above:
\begin{align}
w(\eta) = \frac{1}{2 a}\, \left[\theta(\eta+a)-\theta(\eta-a)\right] .
\label{eqw}
\end{align}
This is a natural choice in a wealth of Monte Carlo simulations and makes explicit analytical solutions for the eigenvalues/eigenfunctions of the Master equation for $a \ge 2L/3$ possible. This will allow us to establish the subtle properties of the eigenspectrum of the Metropolis-Hastings algorithm encoded in Eq.~(\ref{fp.1}), with a fully analytical description of a localisation transition in the relaxation properties. Our results will also demonstrate that for $a = 2 L$, the form Eq.~(\ref{eqw}) for $w(\eta)$ realises the lower bound $\Lambda = 1/2$ in Eq.~(\ref{uniformlimit}) and is therefore an optimal choice. 

In the following, positions and lengths will be measured in units of the box size, so that
$L = 1$. Given the
jump distribution Eq.~(\ref{eqw}), 
the Master equation \eqref{fp.1} reads
\begin{equation}
    P_n(x) \,=\, \frac{1}{2 a}\int_{{\rm max}(-1, x -a)}^{{\rm min}(1, x+a)} P_{n-1}(x') dx' + \frac{|a - 1 + x| +  |a - 1 - x| + 2 a - 2}{4a} P_{n-1}(x).
    \label{eq:fp.2}
\end{equation}
The corresponding eigenvalue/eigenfunction equation for the Master equation then becomes:
\begin{align}
\frac{1}{2 a}\int_{{\rm max}(-1, x -a)}^{{\rm min}(1, x+a)} \Psi(x') dx' + \frac{|a - 1 + x| +  |a - 1 - x| + 2 a - 2}{4a} \Psi(x) = \lambda \Psi(x) ,
\label{eqEvalEvec}
\end{align}
where $\lambda$ is the eigenvalue and $\Psi$ is the corresponding eigenfunction. In the remainder, we use the compact notation $\Psi(x)$ for the eigenfunction, leaving the notation ${\cal P}_\lambda$ only when the full probability distribution $P_n(x)$ is needed. Our analytical method to find the eigenvalue/eigenfunctions
bears some similarities with that used in Ref. \cite{AntalRedner2006}. By differentiating Eq.~(\ref{eqEvalEvec}) with respect to $x$, this integral equation can be turned into a system of coupled differential equations. The number of coupled equations of this system will be found to depend on the jump distance $a$.

\section{Resolution for large jumps ($a>1$)}
\label{sec:large}

\subsection{The case $a \ge 2$: infinite spectral degeneracy}
\label{sec:age2}

For $a > 2$, the eigenvalue equation  Eq.~(\ref{eqEvalEvec}) simplifies into: 
\begin{align}
    \lambda \Psi(x) = \frac{1}{2 a} \int_{-1}^{1} \Psi(y) dy + \left(1 - \frac{1}{a}\right) \Psi(x) .
    \label{eqPsiU}
\end{align}
Integrating over the $(-1,1)$ interval, we find $\lambda S = S$ where $S = \int_{-1}^{1} \Psi(y) dy$. Two solutions are then possible: $S \ne 0$, giving $\lambda = 1$ which corresponds to the steady state distribution ($\Psi = \mathrm{const}$), or $S = 0$, which inserted into Eq.~(\ref{eqPsiU}) gives:
\begin{align}
\lambda_1 = 1 - \frac{1}{a} , \;\; ({\rm for}\; a > 2) .
\label{lambda1P2}
\end{align}
The eigenvalue \eqref{lambda1P2} is infinitely degenerate since all the functions obeying $\int_{-1}^{1} \Psi(y) dy = 0$ are actually eigenvectors. 
Besides, there are no other eigenvalues. For $a = 2$, we realize the optimum, $\lambda_1 = 1/2$, see Eq. \eqref{uniformlimit}.


In this case $a>2$, the full integral equation \eqref{eq:fp.2} reads
\begin{equation}
    P_{n+1}(x) \,=\, \frac{1}{a} \,\frac{1}{2} \,+\, 
    \left(1-  \frac{1}{a} \right) P_n(x) ,
 \label{eq:fp.3}
\end{equation}
taking into account the normalisation $\int_{-1}^1 P_{n-1}(x') dx'=1$.
Equation \eqref{eq:fp.3} admits the following physical interpretation: from any  position $x_0$,
when a jump is attempted, it is accepted with a probability $2L/(2a)=1/a$ since $L=1$,
irrespective of $x_0$. The walker's density after this jump is thus uniform in $[-1,1]$.
With the complementary probability $1-1/a$, the jump is rejected,
in which case $P_{n-1}$ does not change. Rewriting Eq. \eqref{eq:fp.3} as
\begin{equation}
    P_{n}(x)
    \,=\,  \frac{1}{2} \,+\,  \left(1-  \frac{1}{a} \right) \left( P_{n-1}(x)-\frac{1}{2} \right)
 \label{eq:fp.4}
\end{equation}
leads to the solution for all $n$
\begin{align}
P_n(x) = \frac{1}{2} + \left(1 - \frac{1}{a}\right)^n \left( P_0(x) - \frac{1}{2} \right) ,
\label{eq:exact_a_sup_2}
\end{align}
which evidences the eigenvalue $\lambda_1 = 1 - 1/a$.
This behavior is specific to the high symmetry of the uniform potential with a uniform jump distribution and is not representative of more general choices for the confining potential $U(x)$.

While the solution for $a > 2$ is straightforward, the problem for $a < 2$ is more difficult to tackle. We managed to find the analytical solutions for $a \ge 2/3$; this allows to see how the infinitely degenerate eigenvalue $\lambda_1$ evolves under smooth changes of the operator
underlying Eq.~(\ref{fp.1}). 

\subsection{Solution for $1 \le a < 2$}
\label{sec:a1to2}

For $a \le 2$, the eigenvalue equation for initial positions near the walls at $\pm 1$  becomes:
\begin{align}
\lambda \Psi(x) = \left\{
\begin{array}{ll}
\frac{1}{2 a} \int_{x-a}^{1} \Psi(y) dy + \left(\frac{1}{2} - \frac{1 - x}{2 a}\right) \Psi(x) & (x > -1 +a)  \\
\frac{1}{2 a} \int_{-1}^{x+a} \Psi(y) dy + \left(\frac{1}{2} - \frac{x + 1}{2 a}\right) \Psi(x) & (x < 1 - a) . 
\end{array}
\right. \label{Gminus}
\end{align}
For $1 < a \le 2$,  \eqref{eqPsiU} applies 
in the central region $x \in (1 - a, -1 + a)$.
For ``wavefunctions'' describing relaxation modes with $\lambda < 1$, $\int_{-1}^{1} \Psi(y) dy = 0$ and we recover $\lambda = 1 - 1/a$ provided $\Psi(x)$ is non-zero somewhere in $(1 - a, -1 + a)$. The corresponding eigenfunctions  identically vanish for $x < 1 - a$ and $x > -1 + a$ and obey $\int_{-1}^{1} \Psi(y) dy = 0$. The eigenvalue   $\lambda = 1 - 1/a$ thus remains infinitely degenerate. The remaining eigenvalues can be found assuming $\Psi(x) = 0$ for $x \in (1 - a, -1 + a)$ and by transforming the integral equations Eq.~(\ref{Gminus}) into a system of differential equations. 

We introduce
\begin{align}
&\Psi(x) = \Psi_L(1-a-x) \qquad {\rm for} \quad x \in [-1,1 - a] \\
&\Psi(x) = \Psi_R(1-x) \qquad {\rm for} \quad  x \in [-1 + a,1] ,
\end{align} 
so that the arguments of both $\Psi_L$ and  $\Psi_R$ are in 
$(0,2-a)$.
This change of variable allows us to transform the integral equations into two coupled differential equations:
\begin{align}
&\frac{d}{dx} \left[(x-2-2 a(\lambda - 1)) \Psi_L \right] = \Psi_R 
\label{eq:psiR}\\
&\frac{d}{dx} \left[(x + 2 a \lambda - a) \Psi_R \right] = \Psi_L
\label{eq:psiL}
\end{align}
with boundary conditions $\Psi_L(0)=0$. 
The solution reads
\begin{align}
  \Psi_L(x) =  \frac{x (2-3 a +4 a  \lambda )}{2 (1+a  (-1+\lambda )) (-2+x+2 a -2 a  \lambda )} 
+\log\left|\frac{(-2+x+2 a -2 a \lambda)(-a + 2 a \lambda)}{(-2 + 2 a  - 2 a \lambda)(x-a +2 a  \lambda)}\right| .
\label{PsiLeq}
\end{align}
Knowing $\Psi_L(x)$, one obtains $\Psi_R(x)$ from Eq. \eqref{eq:psiR}.

\begin{figure}[h]
\centerline{\includegraphics[clip=true,width=10cm]{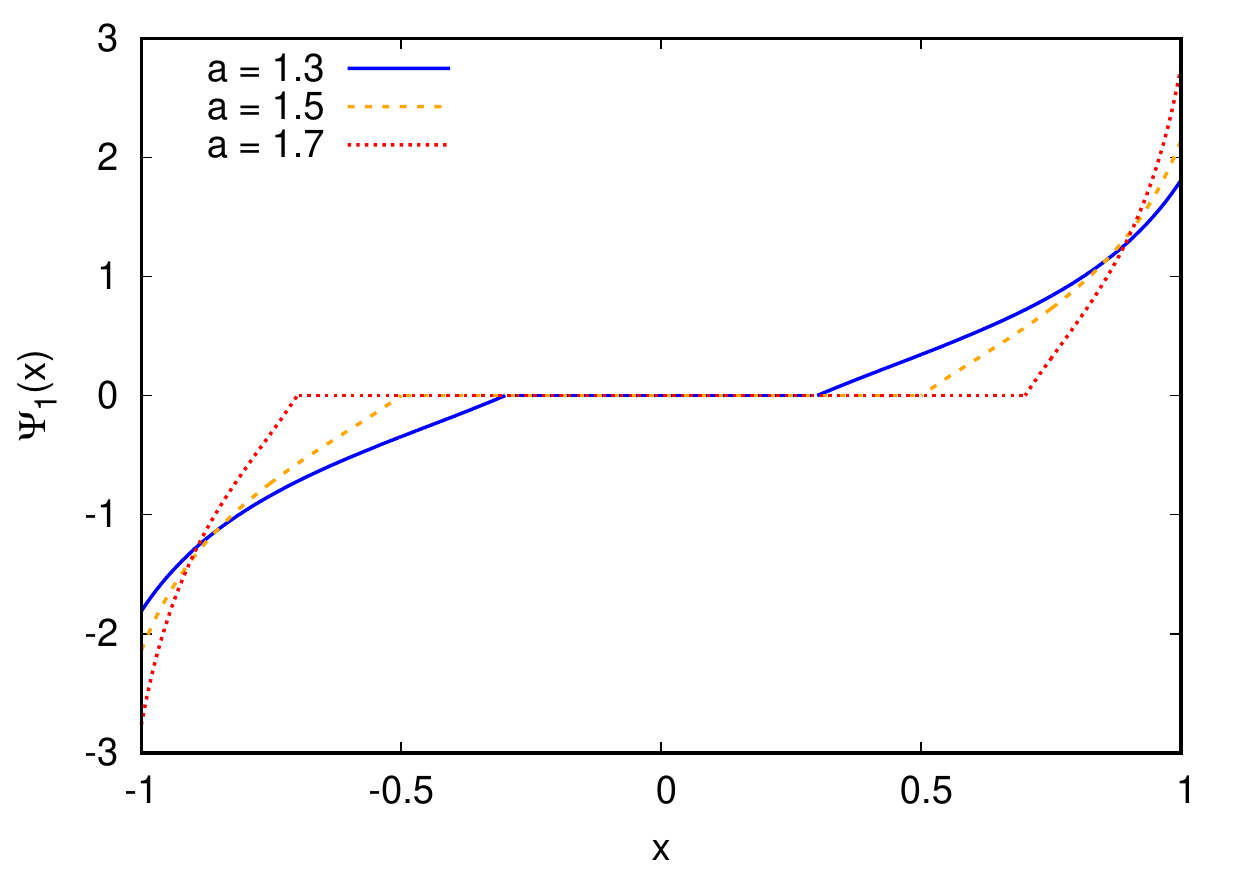}}
\caption{Plot of the eigenfunction $\Psi_1(x)$ (antisymmetric) 
for three values of $a$ in the range $1<a<2$. }
\label{fig:Psi1}
\end{figure}

At this point, the only unknown is $\lambda$, which can be determined from the consistency requirement $\int \Psi(y) dy =0$, which reads
\begin{align}
\int_0^{2-a} dx ( \Psi_L(x) + \Psi_R(x) ) = 0 .
\end{align}
This gives the algebraic eigenvalue equation:
\begin{align}
    -2+3 a -4 a  \lambda + (2 + 2 a \lambda - 2 a) \log\left| \frac{2 + 2 a \lambda - 2 a}{a - 2 a \lambda} \right| = 0 .
    \label{eval1}
\end{align}
We find that for $1 \le a < 2$, this equation has two solutions in the interval $\lambda \in [-1,1]$. One is spurious,  $\lambda = \frac{3 a - 2}{4 a}$: the corresponding ``wavefunction'' $\Psi_L$ from Eq.~(\ref{PsiLeq}) identically vanishes $\Psi_L = 0$. Thus there is only one valid solution; it can be found explicitly rewriting Eq.~(\ref{eval1}) using the variables $(X,Y)$ defined as $\lambda = \frac{-4-X+3 Y}{4(Y-2)}$ and $a = 2 - Y$ in which Eq.~(\ref{eval1}) takes the much simpler form $(X+Y) \log[(X+Y)/(X-Y)] = 2 X$. This allows to find $\lambda_1$ explicitly: 
\begin{align}
\lambda_1 = \frac{(3 - \nu)a +  2 \nu - 2}{4 a}
    \label{lambda1P1}
\end{align}
where $\nu \simeq 1.77187$ is the solution of:
\begin{align}
-\nu + \frac{1+\nu}{2} \log \frac{\nu+1}{\nu-1} = 0 .
\end{align}
It can be checked that the eigenfunction corresponding to Eq.~(\ref{lambda1P1}) is anti-symmetric $\Psi(x) = -\Psi(-x)$. Since this eigenfunction is associated to eigenvalue
$\lambda_1$, we will henceforth denote it by $\Psi_1(x)$
A plot of $\Psi_1$ is given in Fig. \ref{fig:Psi1}. It is continuous, with 
discontinuous derivatives at $1-a$ and $a-1$. 

Equations~(\ref{lambda1P2},\ref{lambda1P1}) describe analytically the behavior of the relaxation rate of the Metropolis algorithm Eq.~(\ref{fp.1}) for a uniform jump distribution $w(x)$. They give an explicit value for $\lambda_1$ in the region $a \ge 1$ and prove that the minimum $\lambda_1 = 1/2$ is attained only at $a = 2$. They also reveal the unusual structure of the eigenvalues in this problem. For $a \ge 2$, there are only two eigenvalues: $\lambda_0 = 1$ which is associated to a symmetric mode, and an infinitely degenerate eigenvalue $\lambda = 1 - 1/a$ which can correspond to both symmetric and anti-symmetric wave-functions. For $1 \le a < 2$, these two expressions for the eigenvalues remain valid with the same degeneracy, but in addition a new non-degenerate mode described by Eq.~(\ref{lambda1P1}) appears, splitting from the $\lambda = 1 - 1/a$ branch at the optimal value $a = 2$. The corresponding eigenfunction is anti-symmetric and localised at the edge of the system where the rejection probability is maximal. 
A summary of the spectral properties will be presented in section \ref{sec:summary}.
The behavior of eigenfunctions is illustrated in Appendix \ref{app:numerical}.

\subsection{Relaxation of symmetric initial conditions for $1 < a < 2$} 
\label{sec:subsymPn}

Solving the eigenvalue equation for $1 < a < 2$ in the previous section, we found that there were only two eigenvalues with $\lambda < 1$ in this range. We also showed that the eigenvector corresponding to $\lambda_1$ was antisymmetric. This raises the question of the relaxation dynamics for symmetric observables for which the contribution from this eigenvector vanishes. We will show here that in general, the relaxation of symmetric observables is not determined by $\lambda_2 = 1 - 1/a$ except if $P_0(x)$ vanishes outside the interval $(-1+a,a-1)$ in which case section \ref{sec:a1to2} shows that $P_0(x) - 1/2$ is in the eigenspace of $\lambda_2$. Rather, the relaxation is ruled by the singular continuum.

Restricting our attention to symmetric observables, we can assume that $P_n(x)$ is symmetric. Then the Master equation~(\ref{fp.1}) can be rewritten in a way that depends only on $P_n(x)$ restricted to $x \in (-1, 1 - a)$.
\begin{align}
P_{n+1}(x) = \frac{1}{2 a} - \frac{1}{2 a}\int^{-x - a}_{-1} P_n(y) dy + \left(\frac{1}{2} - \frac{x+1}{2 a} \right) P_n(x) .
\label{eqPnS1}
\end{align}
It is convenient to rescale $x$ by
introducing $x = -1 + (2 - a) {\tilde x}$ where ${\tilde x} \in (0,1)$. Then expressing $P_n(x)$ as
\begin{align}
P_{n}(x) = \frac{1}{2} + \phi_n({\tilde x})
\end{align}
we find that Eq.~(\ref{eqPnS1}) takes the simpler form
\begin{align}
&\phi_{n+1}({\tilde x}) = - {\tilde K} \int_{0}^{1-{\tilde x}} \phi_n({\tilde y}) d{\tilde y} + \left(\frac{1}{2} - {\tilde K} {\tilde x} \right) \phi_n({\tilde x}) 
\label{eqHN}
\end{align}
where ${\tilde K} = (2 - a)/2a$.

In section \ref{sec:a1to2}, we showed that the eigenvector assumption $\phi_{n}({\tilde x}) = \lambda^n \phi({\tilde x})$ does not lead to solutions with well defined eigenvectors. Thus instead we propose the following anzats for the asymptotic behavior of $\phi_n({\tilde x})$ in the Master equation~(\ref{eqHN}):
\begin{align}
\phi_n({\tilde x}) &= 2^{-n} \left[ e^{-n f({\tilde x})} + h_n \left( e^{-n f(1 - {\tilde x})} - 1 \right) \right]
\label{eqHNanzat}
\end{align}
where the function $f({\tilde x})$ is such that $f(0) = 0$ but $e^{-n f({\tilde x})}$ decays rapidly for $x > 0$. The constant $h_n$ is chosen to ensure $\int_0^1 \phi_n({\tilde x}) d{\tilde x} = 0$. 
This form for $\phi_n({\tilde x})$ allows to replace the integral in Eq.~(\ref{eqHN}) by boundary terms which appear as the dominant asymptotic contribution to the integrals in the limit $n \rightarrow \infty$ by integration by parts. 
For example, we resort to the following approximation : 
\begin{align}
&\int_0^{1-{\tilde x}} e^{-n f({\tilde y})} d{\tilde y} = \int_0^{1-{\tilde x}} \frac{-1}{n f'({\tilde y})} (e^{-n f({\tilde y})})' d{\tilde y} = \frac{1}{n f'(0)} + O(n^{-2})
\end{align}
where we also neglected the second boundary term proportional to $e^{-n f(1- {\tilde x})}$ as it decays exponentially fast with $n$ for ${\tilde x} < 1$.
Using this and similar expansion, together with neglecting all terms which decay faster than $n^{-1}$, we find 
\begin{align}
f({\tilde x}) = -\log(1 - 2 {\tilde K} {\tilde x}) \;,\; h_n = \frac{1}{2 n {\tilde K}} .
\end{align}
Rewriting this result in the original variables, we find the leading asymptotic of the deviation of $P_n(x)$ from its equilibrium value: 
\begin{align}
P_n(x) - 1/2 \propto \left[ R(x)^n - \frac{a}{(2 - a) n} \left( 2^{-n} - R(|x|-a)^n  \right)    \right] \theta(|x|+1-a).
\label{eqLimitSym}
\end{align}
We note that near an edge (for example near the left edge $x = -1$), $\delta P_n(x)$ relaxes to a fixed asymptotic form:  
\begin{align}
\delta P_n(x) \propto \exp(-n (1+x)/ a) .
\label{eqLimitSymScaling}
\end{align}
Interestingly, this dependence resembles the scaling solutions obtained for zero-temperature Monte Carlo dynamics \cite{Chepelianskii_2021},
where the proportionality constant depends on the initial conditions. This form holds if $P_0(x)$ is not identically zero outside of $x \in (1-a,a-1)$. The validity of Eq.~(\ref{eqLimitSym}) is confirmed by direct numerical iteration of the Master equation~(\ref{fp.1}), as shown in Fig.~(\ref{relPsi1})
in Appendix \ref{app:numerical}. We see that for $1 < a \le 2$, the symmetric component of the deviation of $P_n(x)$ from its steady state value $P_\infty(x) = 1/2$ becomes peaked around $x = 1$ 
where $R(x)$ is maximal, but does not exactly tend to a $\delta$ function either because the integral over the entire interval must vanish. 

Generalizing the asymptotic behavior of Eq.~(\ref{eqLimitSym}), we can characterize the relaxation to steady state in this regime by: 
\begin{align}
\lim_{n \rightarrow \infty} \frac{ P_n(x') - P_\infty(x') } { \left({\rm max}_{x'} R(x')\right)^{-n} } = 0 
\label{eq:lim31}
\end{align}
everywhere except at the points $x'$ where rejection probability $R(x)$ is maximal;
at these points, the limit can be non zero and depends on the initial conditions.
It appears that the above relation \eqref{eq:lim31} particularizes 
Eq. \eqref{eq:eigen_discrete_continuous} in the case where the
discrete summation cancels out from symmetry (inherited from the initial conditions).

\section{Solution for $2/3 \le a < 1$}
\label{sec:a2p3to1}

\subsection{The five intervals splitting}
The approach here bears similarities with that for $1 \le a < 2$, where we transformed the eigenvector equation into differential equations by splitting the accessible domain $[-1,1]$ into sub-intervals with a different choice of origin for the expression of the wavefunction in each sub-interval. When  $2/3 \le a < 1$, it is convenient to split the domain $[-1,1]$ into 5 non-overlapping intervals: $I_{-2}, I_{-1}, I_0, I_1, I_2 = [-1, 1 - 2 a], [1 - 2 a, -1 + a], [-1 + a, 1 - a], [1 - a, -1 + 2 a], [-1 + 2 a, 1]$. In each sub-interval, we write $\Psi(x) = \Phi_n(x - y_n)$ where $y_n$ is the center of the interval $I_n$ and $n \in \{-2,-1,0,1,2\}$. With these notations, Equation~(\ref{eqEvalEvec}) can be transformed into the following system of differential equations:
For $x \in [-1+a, 1-a]$:
\begin{align}
& (2 a \lambda) \frac{d \Phi_{-2}}{dx} = \Phi_0 + \frac{d \left[ (2 a - 1 - x) \Phi_{-2} \right]}{dx} \label{Psim2} \\ 
& (2 a \lambda) \frac{d \Phi_{0}}{dx} = \Phi_2 - \Phi_{-2} \label{Psi0} \\
& (2 a \lambda) \frac{d \Phi_{2}}{dx} = -\Phi_0 + \frac{d \left[ (2 a - 1 + x) \Phi_{2} \right]}{dx} \label{Psi2} .
\end{align}
For $x \in [-a_2, a_2]$, introducing $a_2 = 3 a/2 - 1$, we get
\begin{align}
& (2 a \lambda) \frac{d \Phi_{-1}}{dx} = \Phi_1 + \frac{d \left[ (a_2 - x) \Phi_{-1} \right]}{dx} \label{Psim1} \\
  & (2 a \lambda) \frac{d \Phi_1}{dx} = -\Phi_{-1} + \frac{d \left[ (x + a_2)  \Phi_1 \right]}{dx} \label{Psi1} .
\end{align}
The system of equations~(\ref{Psim2},\ref{Psi0},\ref{Psi2}) can be reduced to a second order differential equation on $\Phi_0$: 
\begin{align}
\frac{h_0}{2 x_\lambda} \frac{d}{dx} \left[ (x_\lambda^2 - x^2) \frac{d \Phi_{0}}{dx} \right] + \Phi_0 = A 
\label{eqPsiO}
\end{align}
where $h_{0} = 2 a \lambda, \; x_\lambda = 2 a \lambda - 2 a + 1$ and $A$ is an integration constant, yielding an offset to $\Phi_0$.

Equation~(\ref{eqPsiO}) has the form of a Legendre differential equation and is solved in terms of Legendre functions of the first kind \cite{Polianin}. 
Separating symmetric and antisymmetric components (which are defined up to an overall proportionality constant), we find:
\begin{align}
\Phi_{0S}(x) &= P_\sigma(x/x_\lambda) + P_\sigma(-x/x_\lambda) \label{eqPsi0S} \\
\Phi_{0A}(x) &= P_\sigma(x/x_\lambda) - P_\sigma(-x/x_\lambda) 
\label{eqPsi0A} 
\end{align}
where $P_\sigma$ is the Legendre function with index $\sigma$ given by:
\begin{align}
\sigma &= -\frac{a \lambda + \sqrt{ a \lambda (4 - 8 a + 9 a \lambda)}}{2 a \lambda} \;,\; x_\lambda = 1+2a(-1+\lambda) 
\label{eqSigma}
\end{align}
and the notations $\Phi_{0S}$ and $\Phi_{0A}$ emphasize that these are symmetric/antisymmetric wave-functions respectively.
The wavefunction restricted to the intervals $I_{-2}$ and $I_2$ can be expressed as a function of $\Phi_0$:
\begin{align}
    \Phi_{2}(x) =  &\frac{h_0(x+x_\lambda )}{2 x_\lambda} \frac{d \Phi_{0}}{dx} + A \label{eqPsi2} \\
  \Phi_{-2}(x) = & \frac{h_0 (x-x_\lambda)}{2 x_\lambda} \frac{d \Phi_{0}}{dx} + A .
\label{EQS1} 
\end{align}
We note that $A$ is an overall shift to all the wavefunctions $\Phi_{-2}, \Phi_{0}, \Phi_{2}$.

The solution of the system (\ref{Psim1},\ref{Psi1}) can be written in terms of elementary functions:
\begin{align}
     r_\lambda(x) = \frac{1}{2 w_\lambda (x - w_\lambda)} + \frac{1}{4 w_\lambda^2} \log \frac{w_\lambda - x}{w_\lambda + x} ,
     \label{eqRLdef}
\end{align}
where we introduced the notation $w_\lambda = 2 a \lambda - a_2 = 2 a\lambda - \frac{3 a}{2} + 1$. This leads to
\begin{align}
\Phi_{1}(x) &= C_2 + C_1 r_\lambda(x) \\
\Phi_{-1}(x) &= C_2 - C_1 r_\lambda(-x),
\end{align}
$C_1$ and $C_2$ being two integration constants.

The eigenvalue equations are obtained by requiring the continuity of the wavefunction at the junction between the different intervals. Since we have separated solutions in symmetric/antisymmetric classes, it is enough to write the continuity condition at the junction between intervals $I_0$, $I_1$ and $I_2$. This yields $\Phi_0(1 - a) = \Phi_1(-a_2)$ and $\Phi_1(a_2) = \Phi_{2}(-1 + a)$ (we remind that  $a_2 = \frac{3 a}{2} - 1$).

\begin{figure}[h]
\centerline{\includegraphics[clip=true,width=12cm]{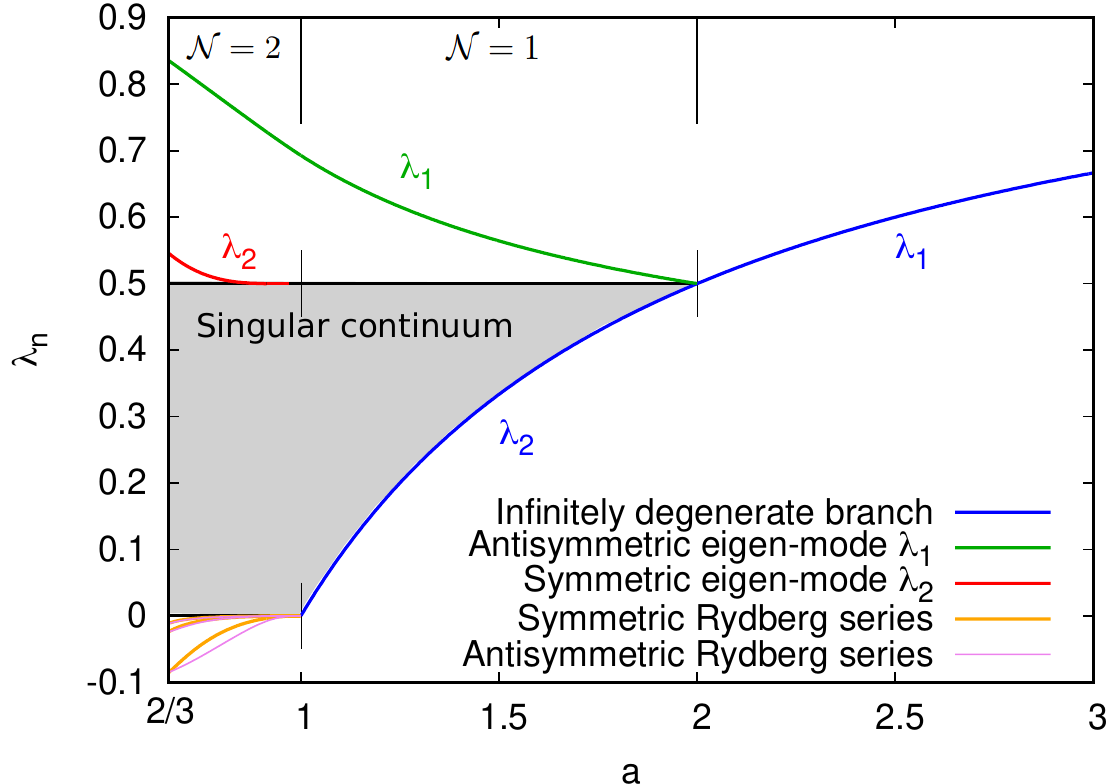}}
\caption{Summary of the exact eigenvalue spectrum $\lambda_n$ ($n \ge 1$) of the Metropolis algorithm Eq.~(\ref{eqEvalEvec}) for $a \ge 2/3$. The upper part of the graph displays the value $\cal N$ of the relevant discrete relaxation modes, see Eq. \eqref{eq:eigen_discrete_continuous}. Besides,
Fig. \ref{figRydberg} corresponds to a zoom into the lower left corner of the present graph,
in the vicinity of the point $a=1$, $\lambda=0$. For $a>2$, the branch shown with eigenvalue $\lambda_1$ is infinitely degenerate.
}
\label{figEvalExact}
\end{figure}

{\em For antisymmetric wavefunctions}, the symmetry fixes $A = C_2 = 0$ and we get
\begin{align}
    r_\lambda(-a_2) \Phi_2(-1 + a) - \Phi_0(1-a) r_\lambda(a_2) = 0 .
    \label{eqLambdaA}
\end{align}
Taking into account Eq.~(\ref{eqPsi0A}) which defines $\Phi_0(x) = \Phi_{0A}(x)$ and  Eq.(\ref{eqPsi2}) giving $\Phi_2(x)$ as a function of $\Phi_0(x)$, we have an explicit equation for the eigenvalue $\lambda$ with anti-symmetric wavefunctions. We denote this eigenvalue by $\lambda_1$ in Fig. \ref{figEvalExact}.

\begin{figure}[h]
\centerline{\includegraphics[clip=true,width=10cm]{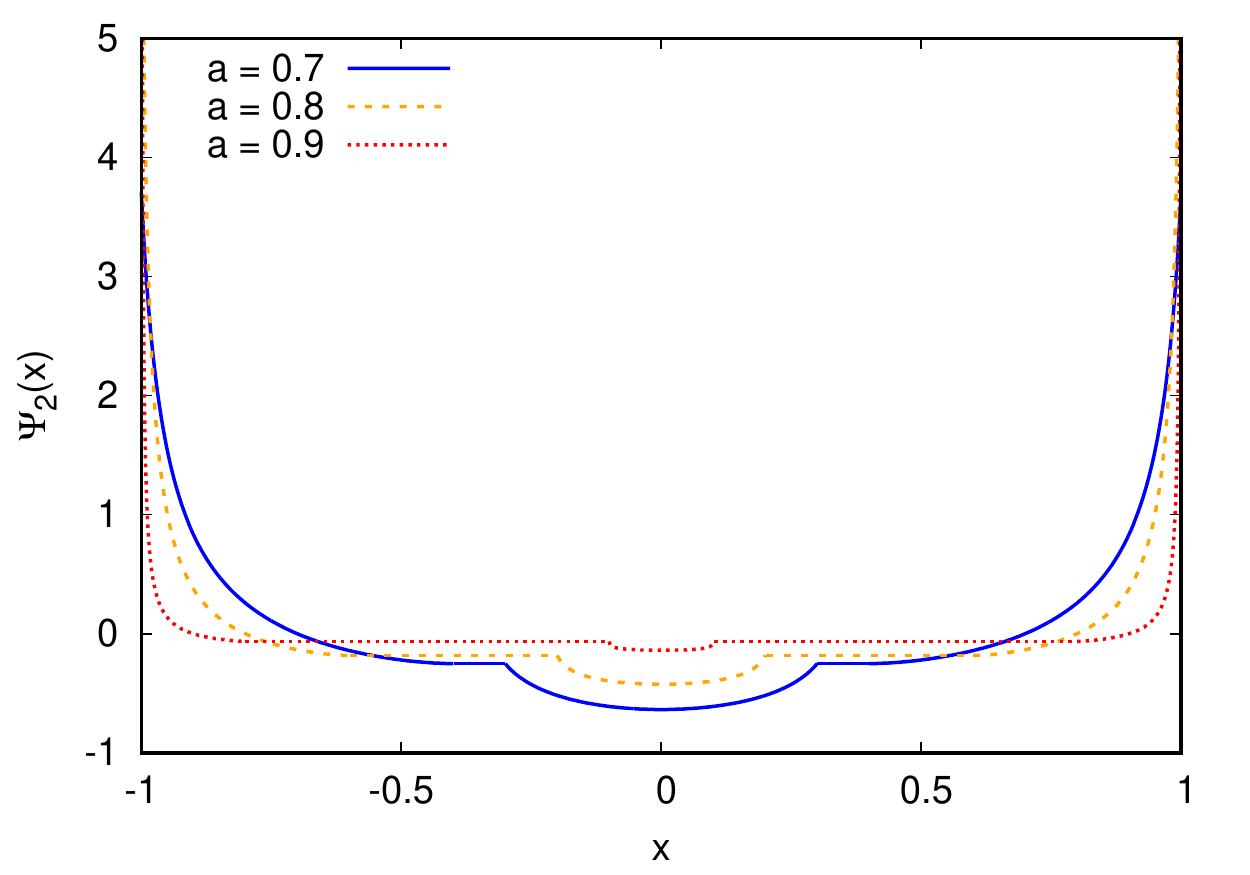}}
\caption{Plot of the eigenfunction $\Psi_2(x)$ 
for three values of $a$ in the range $2/3<a<1$. This function is symmetric,
continuous across the five sectors, with discontinuous derivative at 
the junctions between two consecutive sectors. For $a = 0.8$ and $a = 0.9$, the eigenfunction $\Psi_2(x)$ grows very quickly near the edges of the box $x = \pm 1$ and its maximal value lies outside the chosen vertical range.}
\label{fig:Psi2}
\end{figure}

{\em Within the symmetric subspace}, we have $\Phi_{-1}(-x) = \Phi_1(x)$. This requirement gives $C_1 = 0$ and we find $\Phi_{-1}(x) = \Phi_{1}(x) = C_2 = {\rm const}$. In this case, $A$ is an overall shift of the symmetric solution - and is fixed by $\int_{-1}^{1} \Psi(x) dx = 0$. The matching conditions do not depend on $A$ and we find:
\begin{align}
\Phi_2(-1 + a) - \Phi_0(1-a) = 0
\label{eqLambdaS}
\end{align}
where using Eq.~(\ref{eqPsi0S}) we set $\Phi_0(x) = \Phi_{0S}(x)$ and use Eq.(\ref{eqPsi2}) to find $\Phi_2(x)$ as a function of $\Phi_0(x)$. 
Solving Eq. \eqref{eqLambdaS}, we obtain the eigenvalue $\lambda_2$, shown in Fig. \ref{figEvalExact}.
From the five interval splitting, we reconstruct the whole 
symmetric function $\Psi(x)$, that we denote $\Psi_2(x)$ since it
is associated to $\lambda_2$. It is shown in Fig.
\ref{fig:Psi2}.

The eigenvalue equations Eqs.~(\ref{eqLambdaA},\ref{eqLambdaS}) cannot be solved analytically in their general form. However, they allow to find the eigenvalues numerically and to derive some of their properties. 
For both symmetric and anti-symmetric eigenfunctions, we find an infinite series of negative eigenvalues with the property $\lambda \rightarrow 0$ for $a \rightarrow 1$ and a single positive eigenvalue for both symmetries. These two eigenvalues split away from the rest of the eigenspectrum and are the only eingenvalues above the singular continuum, showing that this case corresponds to ${\cal N} = 2$ . A picture summarizing all the obtained eigenvalues is shown in the next subsection (see Fig.~(\ref{figEvalExact})), while we focus here on the presentation of the analytical results. 

\subsection{Zooming around the $a=1$, $\lambda=0$ point}

\begin{figure}[h]
\centerline{\includegraphics[clip=true,width=12cm]{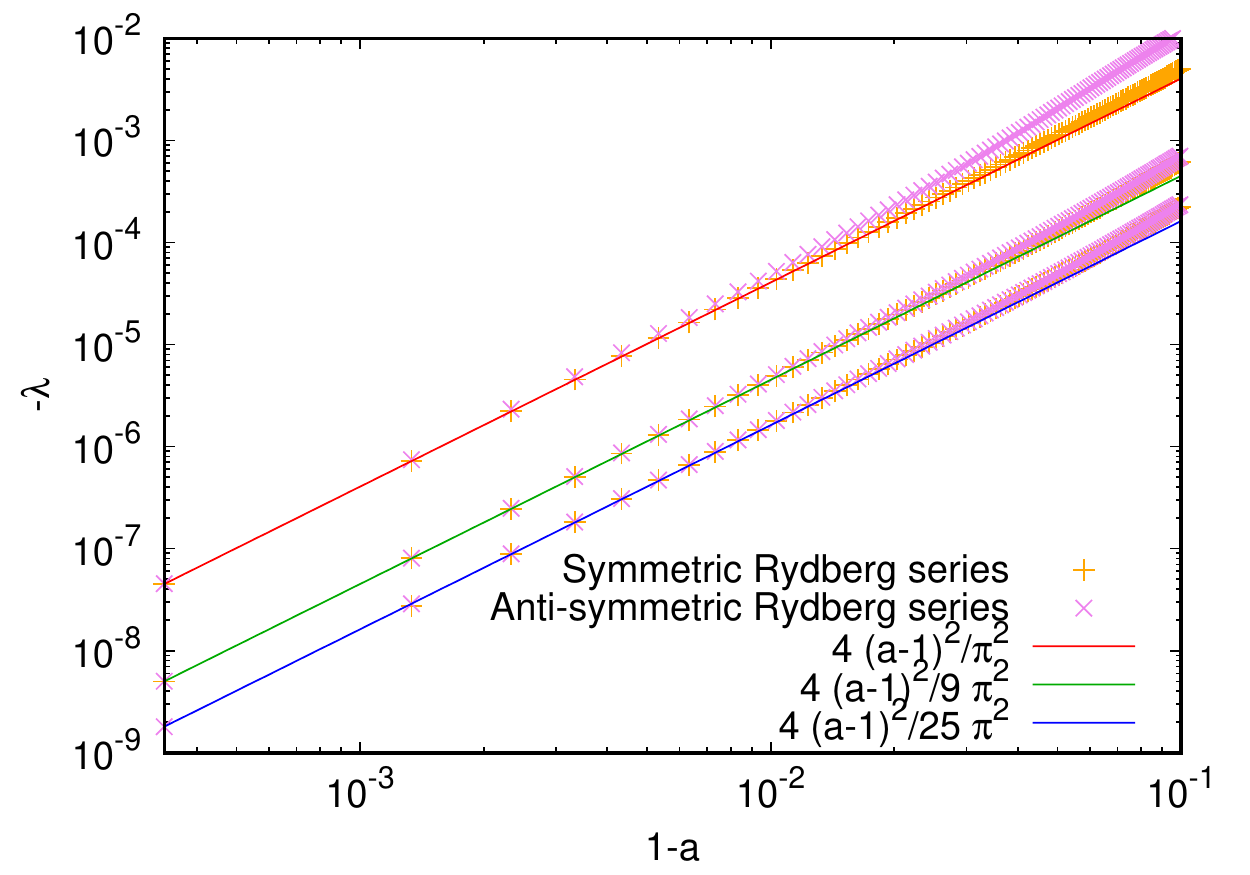}}
\caption{Scaling behavior of negative eigenvalues for $a < 1$ in the limit $a \rightarrow 1$.
The first three negative eigenvalues with the largest modulus are plotted for symmetric/anti-symmetric wave functions
(for $n=0,1$ and 2). In the limit $a \rightarrow 1$, even/odd wavefunctions become degenerate and follow the Rydberg series Eq.~(\ref{eqRydberg}). 
}
\label{figRydberg}
\end{figure}

We turn to investigate the asymptotic properties of the eigenvalues for $\lambda \rightarrow 0$ and $a \rightarrow 1$. From Eq.~(\ref{eqSigma}), this limit corresponds to $\sigma \simeq \frac{1}{\sqrt{-\lambda}} \rightarrow \infty$ and  $\frac{1-a}{x_\lambda} \simeq a -1 \rightarrow 0$. Analysing the asymptotic properties of Eqs.~(\ref{eqLambdaS},\ref{eqLambdaA}) in this limit, we find a Rydberg-like series of eigenvalues (with odd denominators):
\begin{align}
    \lambda^{(s,a)}_n \simeq -\frac{4(a-1)^2}{\pi^2(1+2n)^2} .
\label{eqRydberg}
\end{align}
Remarkably, the leading order asymptotic behavior coincides for symmetric $\lambda^{(s)}_n$ and anti-symmetric $\lambda^{(a)}_n$ wavefunctions.
The asymptotic Rydberg series behavior of the negative eigenvalues is compared with the numerical eigenvalues on Fig.~(\ref{figRydberg}), which illustrates this feature.
When $a$ departs from unity, the symmetric and anti-symmetric branches of eigenvalues move further away from each other.

The anti-symmetric eigenvalue $\lambda_1$ is a smooth continuation of the solution Eq.~(\ref{lambda1P1}) obtained for $a \in (1,2)$ into the interval $a \in (2/3,1)$. We will thus not discuss its analytical properties. The analytical behavior of the symmetric eigenvalue $\lambda_2$ is more subtle, as in the previous section we established that there are no symmetric eigenvalues for $a > 1$. 

Numerically, we find that $\lambda_2$ rapidly tends to a limit value $\lambda_2 = 1/2$ as $a$ approaches unity from below. It is thus natural to expand Eq.~(\ref{eqLambdaS}) in this limit. We found, however, that to obtain accurate asymptotic estimates, a change of variables is needed. We introduce the variable $X = 1 - \frac{1-a}{x_\lambda}$, which is the deviation from unity of the (positive) argument of the Legendre functions in Eq.~(\ref{eqPsi0S}), evaluated at $x=1-a$. Expanding Eq.~(\ref{eqLambdaS}) in the small parameters $X$ and $1-a$ (using $x_\lambda = 2 a \lambda_2 - 2 a + 1$, one  sees that $\lambda \rightarrow 1/2$ corresponds to $X \rightarrow 0$). We remind that in this expansion, the parameter $\sigma$ of the Legendre functions has also to be expressed as a function of $X$. Performing this expansion, we get the following Laurent series for $\log X$ as a function of $1-a$:
\begin{align}
\log X = -\frac{1}{2(1-a)} - \frac{3}{2} + \log 2 + (1 - a)(\pi^2/3-4) + ...  
\end{align}
Inserting this solution into $X = 1 - (1-a)/x_\lambda$ and treating it as an equation on $\lambda_2$ (through $x_\lambda$), we find:
\begin{align}
\lambda_2 - \frac{1}{2} = \exp\left( -\frac{1}{2(1-a)} + \log(1-a) - \frac{3}{2} + \frac{(1-a)(\pi^2 - 15)}{3} + ...\right)
\label{eqLambda2expansion}
\end{align}
This asymptotic expansion demonstrates the non analytic behavior of $\lambda_2$ near $1/2$, and we also see that the expansion of $\log(\lambda_2 - 1/2)$ does not take the form of a Laurent series in $a-1$ because of the additional $\log(1-a)$ terms. It is probable that this logarithmic term explains why the direct expansion in the parameters $\lambda_2-1/2$ was not successful. Since the derivation of Eq.~(\ref{eqLambda2expansion}) relies heavily on the use of formal mathematical calculation software, we checked the accuracy of Eq~(\ref{eqLambda2expansion}) numerically to confirm the validity of our derivations. We find that for $a = 0.95$, the relative error between the numerical value of $\lambda_2$ and the estimate Eq~(\ref{eqLambda2expansion}) is $3.9 \times 10^{-3}$, decreasing further as $a$ approaches unity. 

\section{Summary of the exact eigenvalues spectrum}
\label{sec:summary}

In essence,  we obtained a complete description of the eigenvalue spectrum of the Metropolis Master Eq.~(\ref{eqEvalEvec}) in the range $a \ge 2/3$. This provides us with a reference case to understand the structure of the eigenvalue spectrum for this type of self-adjoint operator, which markedly differs from the spectrum of a Schr\"odinger equation in a box. Fig.~(\ref{figEvalExact}) shows the full eigenspectrum for all the range  $a \ge 2/3$,  for which some explicit analytic solutions could be obtained; this conveys a summary of our findings.

{\bf Range $a \ge 2$}: We obtained in section \ref{sec:age2} an explicit analytic solutions for $a \ge 2$. This case corresponds to ${\cal N} = 1$, with the particularity that the eigenvalue $\lambda_1$ is infinitely degenerate. The choice $a = 2$ gives $\Lambda = \lambda_1 = 1/2$, since 1/2 is also the rejection probability at the edge of the box (and thus a lower bound for $\Lambda$); this choice realises the optimal convergence rate for this problem, among all possible choices of a jump distribution function $w(x)$.

{\bf Range $1 \le a < 2$}: This case, treated in Sec. \ref{sec:a1to2},
also corresponds to ${\cal N} = 1$ but the eigenvalue $\lambda_1$ is non degenerate and corresponds to an anti-symmetric eigenfunction. This eigenvalue is also the only eigenvalue below 1, which lies above the singular continuum $R([-1,1])$. By symmetry, any symmetric initial probability distribution $P_0(x)$ has a vanishing matrix overlap with this eigenmode. The relaxation of such a probability distribution will thus be governed by the singular continuum. The deviation of $P_n(x)$ from $P_\infty(x)$ will then be given by localizing term $\delta P_n(x) \simeq {\cal L}_n(x)$, with the property that $\lim_{n \rightarrow \infty} {\cal L}_n(x)/{\cal L}_n(\pm 1) = 0$ for any $x \ne \pm 1$. The points $x = \pm 1$ at the edge of the box are those with the highest rejection probability $R(x)$. The leading asymptotic behavior of ${\cal L}_n$ was established in Sec.~\ref{sec:subsymPn}.

{\bf Range $2/3 \le a < 1$}: In Sec.~ \ref{sec:a2p3to1}, we derived the exact solution to the associated eigenvalue problem. This allowed us to establish that $a = 1$ is a critical point corresponding to the transition from ${\cal N} = 2$ (for $2/3 < a < 1$) to ${\cal N} = 1$ (for $1 < a < 2$). To show this, we derived an algebraic equation on the eigenvalues involving Legendre functions of the first kind (see section \ref{sec:a2p3to1}). These equations could no longer be solved explicitly. We found eigenvalues numerically and derived some of their asymptotic properties in the limit $a \rightarrow 1$, thereby characterizing the transition ${\cal N} = 2 \rightarrow {\cal N} = 1$.

The main result of this paper is the analytical description of the eigenspectrum at the transition between ${\cal N} = 1$ and ${\cal N} = 2$. For $2/3 < a < 1$, a symmetric eigenfunction solution appears which does not exist for $a > 1$; it yields the eigenvalue $\lambda_2$ with a limit $\lambda_2 \rightarrow 1/2$ for $a \rightarrow 1$ described by equation \eqref{eqLambda2expansion}. The eigenvalue $\lambda_2$ merges with the top of the singular continuum $\Lambda = 1/2$ following a dependence which is reminiscent of the gap equation in superconductivity \cite{Kittel} with a leading term $\lambda_2 - 1/2 \sim \exp([2(a-1)]^{-1})$, which cannot be expanded as a power series in $a-1$.
We also found that this change in ${\cal N}$ goes together with the appearance of an (approximate) Rydberg series of negative eigenvalues  ($a < 1$), which branch off from the infinitely degenerate eigenvalue $\lambda_2 = 1 - 1/a$ ($a > 1$). This last observation may be specific to the box problem because of the existence of an infinitely degenerate branch $\lambda_2$ for $1 < a < 2$. On the other hand, we expect that the merging between a relaxation eigenmode and the singular continuum, that we described here, is a generic phenomenon for transitions between relaxation regimes with different number of non-equilibrium relaxation eigenmodes ${\cal N}$ that appear in the relaxation expansion \eqref{eq:eigen_discrete_continuous}.

\section{Conclusion}
\label{sec:conclusion}

We investigated the relaxation properties of a simple Monte Carlo algorithm, describing a random walker confined in a (rescaled) box $x \in (-1,1)$ and subject to a flat potential $U(x) = {\rm const}$. The equilibrium probability distribution is thus uniform in $(-1,1)$. At each algorithmic step, we attempt to change the particle position $x \rightarrow x + \eta$ where the jump distance $\eta$ is drawn from a uniform probability distribution in $(-a, a)$. The length scale $a$ is the maximum attempted jump length. If the attempted jump falls outside the box $(-1,1)$, the move is rejected and $x$ remains unchanged. Otherwise the move is accepted and the position is updated to $x + \eta$. For any $a > 0$, the probability distribution of the random walkers will converge towards equilibrium, with a strongly $a$-dependent relaxation rate.

To analyse the relaxation rate, we studied the Master equation describing this  process. It is an integral equation which in some cases can be mapped to a system of differential equations. For $a > 2/3$, we solved this system analytically to obtain the full spectrum of eigenvalues and eigenfunctions. We found that, in general, the relaxation dynamics is not described only by the eigenvalue spectrum. Instead, we need to take into account the contribution of the set of singular eigenvalues equal to the rejection probability $R(x)$ at some point $x \in (-1,1)$. A possible eigenfunction associated to such a singular eigenvalue is zero everywhere except at the point $x$. This function vanishes almost everywhere and has a vanishing norm. To highlight the unusual nature of such eigenfunctions, we refer to this part of the eigenspectrum $R([-1,1])$ as the singular continuum. This name is meant to distinguish these eigenvalues from the continuum of eigenvalues that can appear for the Schr\"odinger equation on the real line, for which on the contrary, the norm of the eigenvectors becomes infinitely large. In general, the contribution from the singular continuum will dominate the asymptotic convergence to equilibrium only after taking into account the contribution of a number+ ${\cal N}$ of relaxation eigenmodes. We recently showed some examples where a transition to ${\cal N} = 0$ can occur. In these situations, the relaxation behavior is completely dominated by the singular continuum \cite{MC2022}; however, we could not study analytically the transition between different values of ${\cal N}$. 

For the present free particle confined in a box, such a study is possible; this is the main result of our paper. We showed that the regime $1 < a <2$ corresponds to ${\cal N} = 1$, with a single anti-symmetric relaxation eigenmode above the singular continuum. This eigenmode thus has vanishing overlap with symmetric initial probability distributions; in this case, the singular continuum rules the dynamics, and the deviation from equilibrium concentrates at late time (after a large number of iterations) onto the points of maximal rejection, at the edge of the box $\pm 1$. 
Indeed, the singular continuum eigenfunctions vanish almost everywhere. We then showed that ${\cal N} = 2$ for $2/3 < a < 1$. The transition is described by a merging between a relaxation eigenmode at $2/3 < a < 1$ and the singular continuum which takes place at $a = 1$. We obtained asymptotic estimates describing the gap between the eigenvalue spectrum and the top of the singular continuum. Interestingly, this gap vanishes to all orders in $1-a$ and cannot be obtained by a regular perturbative expansion near $a = 1$. We believe that this merging with the continuum is a generic scenario describing the change in the number of relaxation eigenmodes ${\cal N}$.
At the same point $a=1$, we evidenced the emergence of a Rydberg-like series of negative eigenvalues;
the connection between the two phenomena remains an open question.

\bibliography{mainbib.bib}


\begin{appendix}
\section{Comparison with numerical diagonalization and Monte Carlo simulations} 
\label{app:numerical}

\begin{figure}[h]
\includegraphics[clip=true,width=8cm]{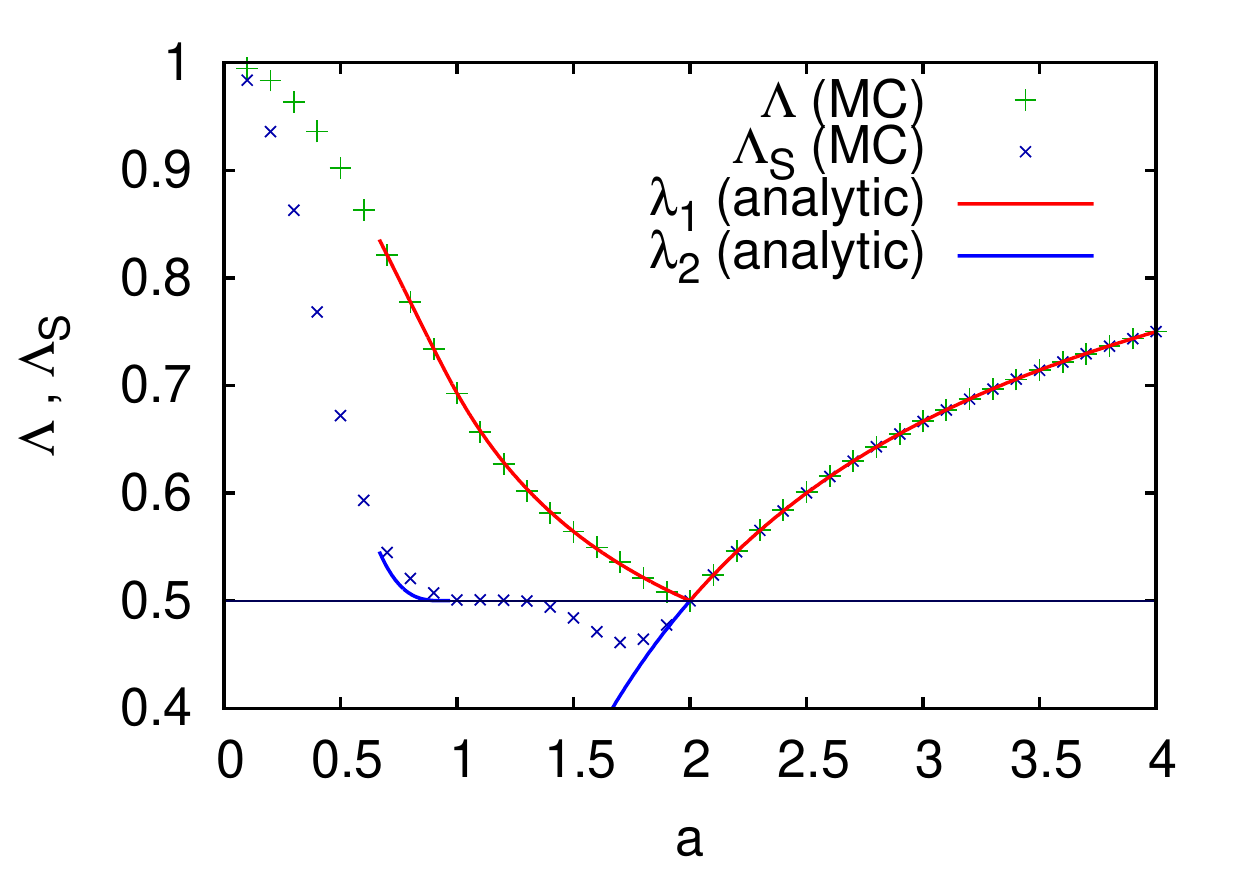}
\caption{Comparison between the relaxation rate obtained numerically from the Monte Carlo dynamics (dots) and our analytical calculations
(continuous lines). 
The Monte Carlo data is collected over $10^{10}$ independent simulations up to $n=50$ with initial conditions $P_0(x) = \delta(x-1)$. The values for $\Lambda$ are extracted from by calculating the mean value of $\mathcal{O}(x) = |x+0.5|$ and for $\Lambda_S$ from a symmetric observable $\mathcal{O}(x) = |x|$ (the subscript $S$ thus refers to the symmetric branch). We attribute the values below $1/2$ for $\Lambda_S$ to incomplete numerical convergence.}
\label{figEvalApprox}
\end{figure}

To put our analytical findings to the test, we performed several numerical simulations: direct simulations of the Monte Carlo algorithms and numerical studies of the Master equation (including diagonalization of discretized versions of the Master equation, or iteration of the Master equation itself, starting from a specific initial probability distribution $P_0(x)$). In Fig.~\ref{figEvalApprox}, we show the agreement between the relaxation rate ${\Lambda}$ estimated from the simulations of the Monte Carlo algorithm. We see a very good agreement between the relaxation rate ${\Lambda}$ obtained for arbitrary (non symmetric) initial conditions and our analytical results. The agreement for the next leading eigenvalue corresponding to symmetric $P_0(x)$ is less accurate. In particular we see some deviations close to the transition ${\cal N} = 2 \rightarrow 1$, we attribute those to more pronounced transients effects close to this transition.

\begin{figure}
\begin{tabular}{cc}
\includegraphics[clip=true,width=8cm]{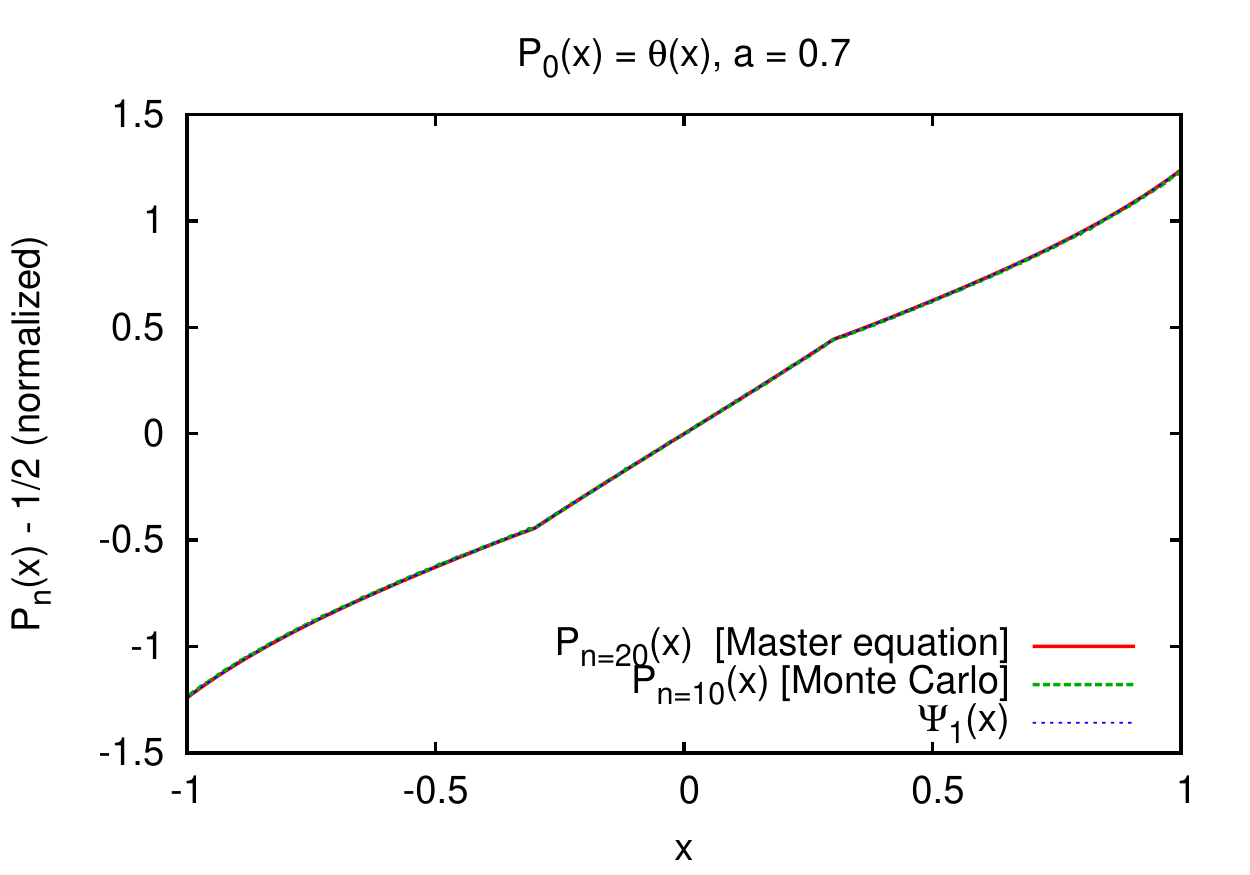}     &  
\includegraphics[clip=true,width=8cm]{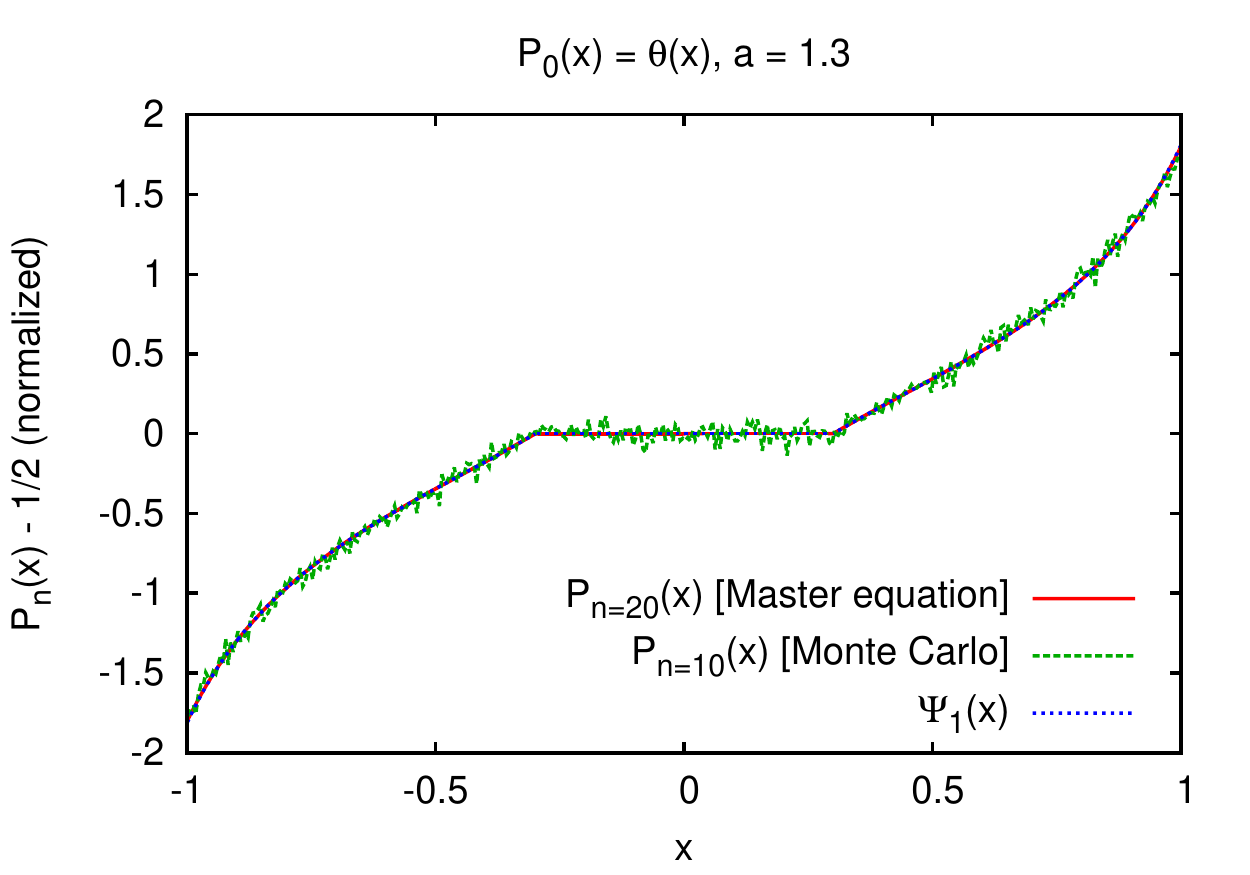} \\
\includegraphics[clip=true,width=8cm]{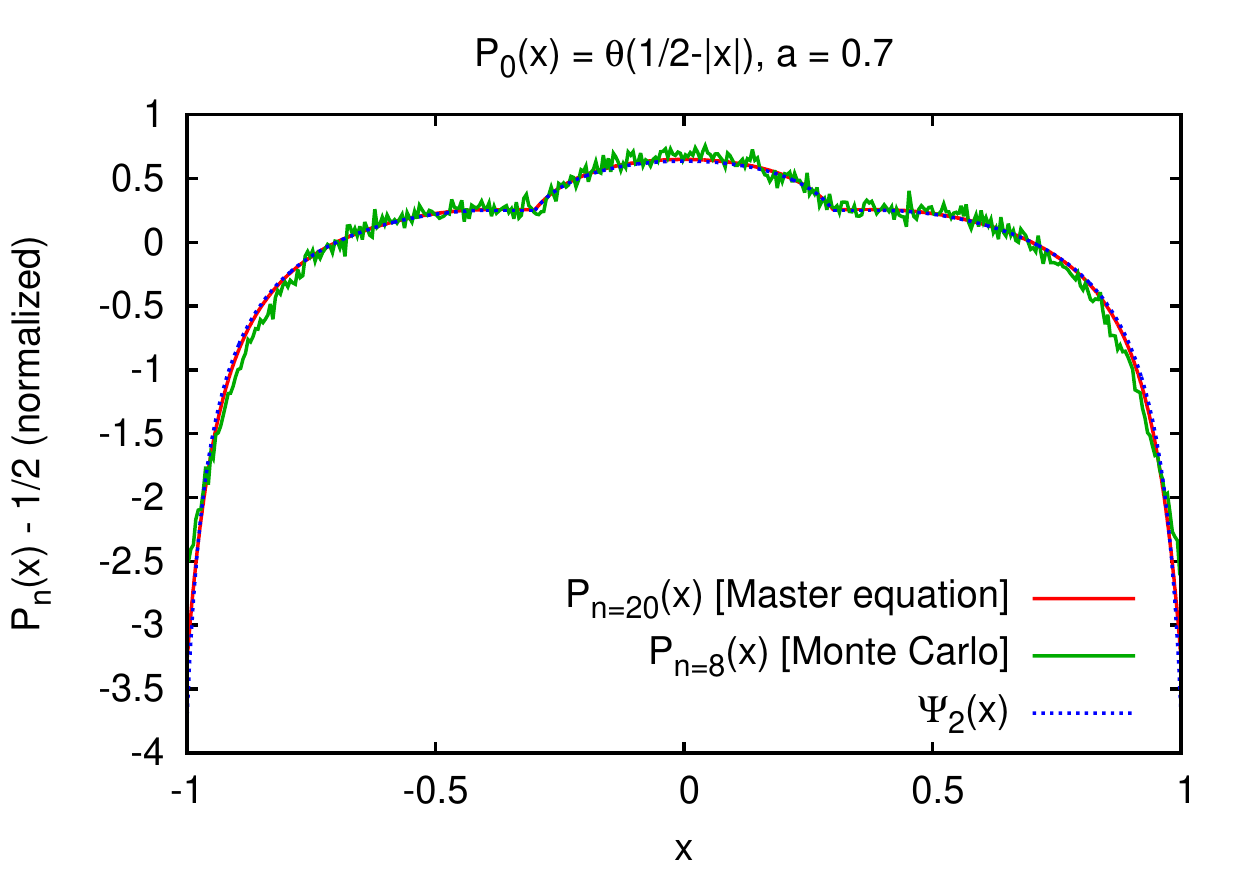}     &  
\includegraphics[clip=true,width=8cm]{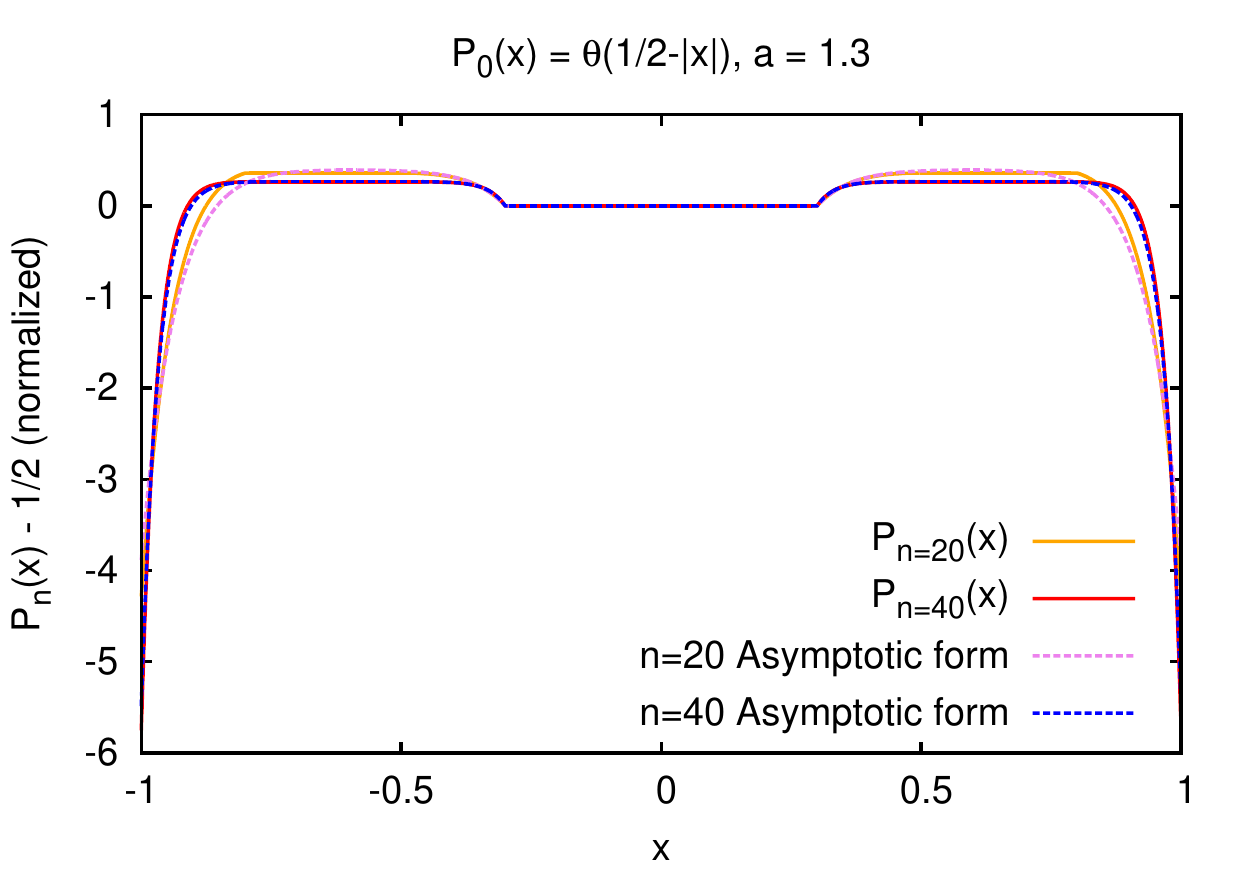} 
\end{tabular}
\caption{The top panels show the relaxation of the deviation from equilibrium $P_n(x)-1/2$ (normalized to unity) towards the eigenvector $\Psi_1(x)$ starting from an asymmetric initial condition for $a = 0.7$ and $a = 1.3$. The bottom panels show the same quantity, starting from symmetric initial conditions for $a < 1$ the normalized $P_n(x)-1/2$ converges to the analytical $\Psi_2(x)$ but for $a > 1$, the deviation from equilibrium becomes more and more localized at the edges as $n$ increases. The discretization used $N = 5000$ points uniformly distributed in the box $(-1,1)$. The Monte Carlo data are collected over $10^{10}$ independent iterations of the dynamics. In the bottom-right panel, comparison is made with the asymptotic prediction
of Eqs. \eqref{PsiLeq} for the case $a>1$ in the upper right panel,
with the results of section \ref{sec:subsymPn} for the bottom right panel; for $a<1$ (upper and bottom left panels), we use the results of section \ref{sec:a2p3to1}.
}
\label{relPsi1}
\end{figure}

The change in the relaxation properties of the Monte Carlo dynamics at $a = 1$ is illustrated on Fig.~\ref{relPsi1}, which displays the (normalized) deviation from equilibrium $\delta P_n(x) = P_n(x) - 1/2$ after $n$ steps of the Monte Carlo algorithm. 
We report results for $a = 0.7$ and $a= 1.3$, for two initial probability distributions: a generic $P_0(x)$ without any symmetry, and a symmetric function $P_0(-x) = P_0(x)$. Numerical results are obtained using both direct Monte Carlo simulations and iteration of the Master equation, except for the symmetric $a = 1.3$ case for which only results from the Master equation are shown, as a larger number of step $n$ was required.
For $a = 0.7$, both choices of $P_0(x)$ relax to the leading relaxation eigenmode, with a symmetry compatible with that of $P_0(x)$: we observe
an anti-symmetric eigenmode for a generic $P_0(x)$, and the sub-leading symmetric eigenmode for a symmetric $P_0(x)$. Relaxation to this eigenmode is fast, requiring about $\sim 10$ steps of the algorithm. For $a = 1.3$, a similar picture is observed for a generic $P_0(x)$, but the situation for symmetric $P_0(x)$ is very different. We see that $\delta P_n$ does not converge to fixed eigenfunctions but instead keeps concentrating at the edges of the box, at $x=\pm 1$, where the rejection probability is highest. This corresponds to the localizing contribution coming from the singular continuum. The numerical results are compared with the predicted asymptotic form for $\delta P_n$, that we derived in Sec.~\ref{sec:subsymPn}, showing a very good agreement. 

\begin{figure}
\begin{tabular}{cc}
\includegraphics[clip=true,width=8cm]{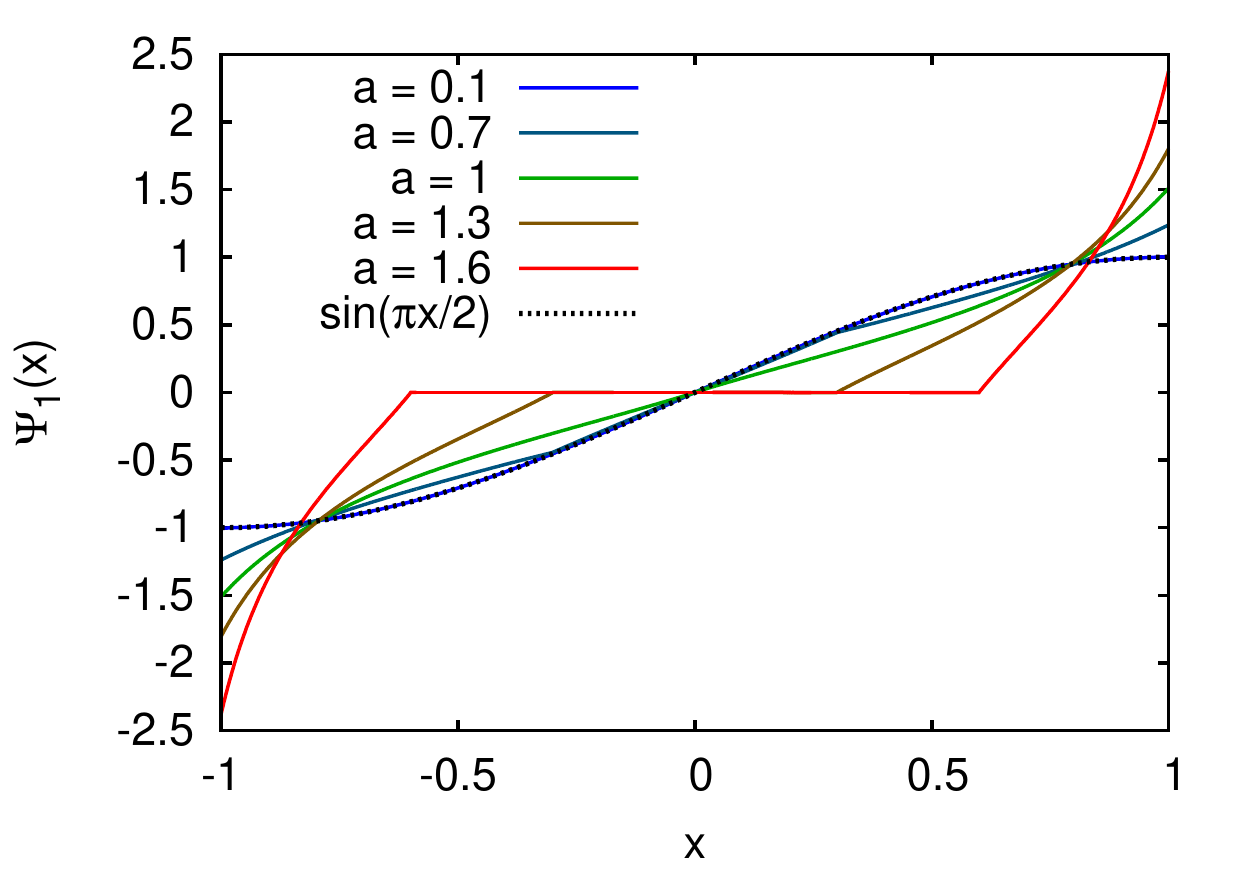}     &  
\includegraphics[clip=true,width=8cm]{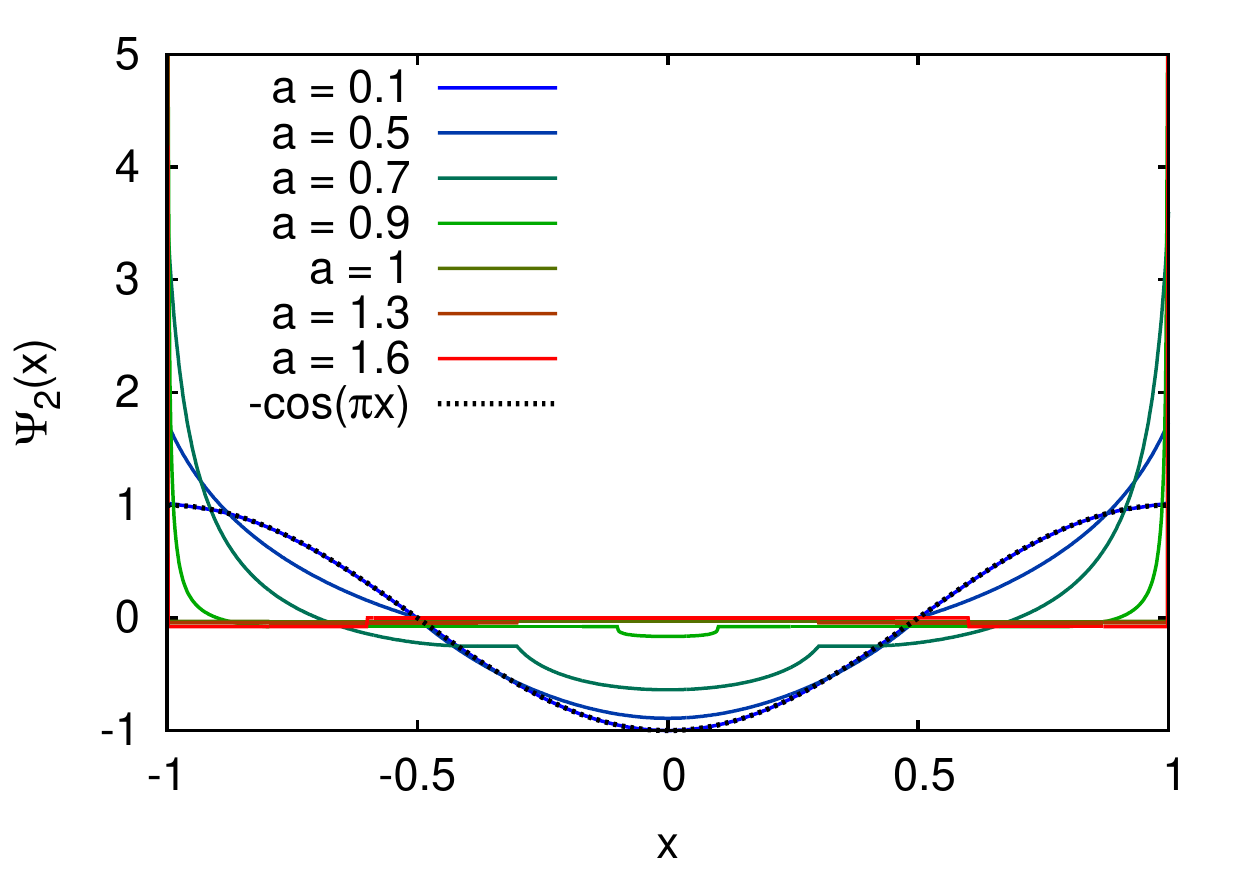} 
\end{tabular}
\caption{Evolution of the numerical eigenvectors $\Psi_1(x)$ and $\Psi_2(x)$ with the jump distance $a$, for $N = 1000$ discretization sites. For $\Psi_2(x)$, most of the weight is localized at the edge points $\pm 1$: $\Psi_2$ falls outside of the vertical range displayed, as a consequence of localization.
}
\label{psiVsA}
\end{figure}

Finally, we illustrate the evolution of the leading relaxation eigenmodes with the (maximal jump length) parameter $a$, see Fig. 
\ref{psiVsA}. At $a \ll 1$, the two leading relaxation eigenmodes are very close to the solutions of the corresponding Fokker-Planck problem: $\Psi_1(x) \simeq \sin \pi x/2$ and $\Psi_2(x) \simeq \cos \pi x$. As $a$ increases, the eigenmodes feature a larger weight on the edges, where the rejection probability is maximum. For $a > 1$, we see that the numerically simulated $\Psi_2$ completely collapses to the edge of the system (localization). This is a manifestation of the absence of symmetric relaxation eigenmodes above the singular continuum, for $1 < a < 2$.

\end{appendix}

\end{document}